\documentclass[%
 reprint,
 amsmath,amssymb,
 aps,nofootinbib,   
]{revtex4-1}
\usepackage{bm}
\PassOptionsToPackage{linktocpage}{hyperref}
\usepackage[hyperindex,breaklinks]{hyperref}
\usepackage{enumitem}
\usepackage{slashed}

\renewcommand{\theta}{\vartheta}

\usepackage{array}
\usepackage{mathtools}

\usepackage{etoolbox}

\begin{document} 

\title{On Exclusion of Positive Cosmological Constant}


\author{Gia Dvali$^{a,b,c}$  and  Cesar Gomez$^{d}$} 
\affiliation{
$^a$Arnold Sommerfeld Center, Ludwig-Maximilians-Universit\"at, Theresienstra{\ss}e 37, 80333 M\"unchen, Germany, 
}
 \affiliation{
$^b$Max-Planck-Institut f\"ur Physik, F\"ohringer Ring 6, 80805 M\"unchen, Germany
}
 \affiliation{ 
$^c$Center for Cosmology and Particle Physics, Department of Physics, New York University, 726 Broadway, New York, NY 10003, USA
}
\affiliation{$^d$Instituto de F\'{\i}sica Te\'orica UAM-CSIC, Universidad Aut\'onoma de Madrid, Cantoblanco, 28049 Madrid, Spain}


\begin{abstract}
Some time ago we have suggested that positive vacuum energy 
  exhibits a finite quantum break time, which can be a signal that 
  a positive cosmological constant is inconsistent.  From the requirement that 
  Universe never undergoes through quantum breaking, we have derived 
  an absolute lower bound on the speed of variation of positive vacuum energy.   The same suggestion about exclusion of positive cosmological constant was made recently. We show that the new bound 
  represents a particular string theoretic version of the old bound, 
  which is more general.  
In this light, we  show that the existing window
  still provides a large room for the inflationary and dark energy model building. In particular, the inflationary models with gravitational strength interactions, are protected against fast quantum breaking. 
  
\end{abstract}


\maketitle
     In a series of works \cite{QB}, we have shown that    
  space-times with positive cosmological constant  exhibit a finite quantum breaking time,  $t_Q = H^{-1}\alpha^{-1}$,  where $H$ is the Hubble parameter in the units of 
  the Planck mass, which we shall use throughout  this note, and 
  $\alpha$ is the quantum interaction strength of constituents. In pure gravity, 
  $\alpha = H^2$ and thus, $t_Q = H^{-3}$.   
  After the time $t_Q$ the mean-field evolution of the Hubble patch 
 can no longer match the solutions of classical gravity.  What replaces the classical evolution after $t_Q$, is unknown. One possibility - that would be especially favoured from the point of view of the {\it Cosmological Constant Problem} - is that the quantum breaking is a signal of inconsistency and hence should never happen.  This requirement imposes a lower bound on the speed 
 of variation of any sort of positive vacuum energy, including the 
 potential energy of scalar fields.  This bound,  written 
 as a constraint on the gradient of the scalar potential  $V$ in any 
 canonically-normalized scalar field direction, reads  
 \begin{equation} \label{central} 
   |\nabla V| \, > \, {\sqrt{V} \over t_Q} = V \alpha \, .
 \end{equation}  
 For a gravitational strength coupling, which represents an absolute lower bound on
 $\alpha$, we have $\alpha = V$, whereas for certain critical systems we can have $\alpha = {\rm ln}^{-1}(V^{-1})$, which give 
   \begin{equation} \label{central1} 
   |\nabla V| \, > \,  V^2 \,, ~{\rm and}~ |\nabla V| \, > \,  V{\rm ln}^{-1}(V^{-1}) 
 \end{equation}  
respectively. 
   
   In a recent paper \cite{new}, a very similar form of the constraint 
   was suggested 
   \begin{equation} \label{new} 
   |\nabla V| \, > \, c V \, ,
 \end{equation}   
where $c$ is restricted to be some $V$-independent constant.  
The former criterion is only informative if some lower bound on $c$ is provided. Unfortunately, to our understanding, the authors cannot provide any consistency lower bound on $c$. Instead, in \cite{new} an upper bound on $c$ is derived from some systematic study of different efforts to construct deSitter in string theory. 
 It is not a purpose of this note to enter into this analysis. 
 However, our interpretation is that the suggested new bound (\ref{new}) 
 is a particular string theoretic version of (\ref{central}) with 
 $\alpha = c$ taken as a constant, in which case 
 we expect that $c$ is bounded from above  by a quantum string coupling, 
 $c \sim g_{str}^2$, which fully matches  (\ref{central}) with 
 $\alpha = g_{str}^2$.  In this sense, the upper value of $c$ observed  in 
 \cite{new}, is just a reflection of the fact that in the examined models 
 the string coupling is within the perturbative limit. 
  
 In order to explore the full power of the quantum breaking bound, we must  abstract from the specifics of string theory 
 inflationary model building.  In this case we must use the more general bound (\ref{central}), rather than (\ref{new}). 
 For example,  In the absence of a lower bound on $c$, the criterion (\ref{new}) cannot serve as  evidence for exclusion of constant vacuum energy.  In contrast, the bound (\ref{central}) does exclude such 
 a possibility, because $\alpha$ is bounded from below by $\alpha=V$.  Then, taking the latter as the correct lower bound, we get a very large window of opportunity for the inflationary model building. 
 For example, any slow-roll model with gravitational-strength interactions, 
 such as Linde's chaotic model, satisfies the bound (\ref{central}).   

 Likewise, the restriction on the time-variation of the present 
 dark energy is extremely mild.  For example,  the bound on the time-scale of variation of the present cosmological ``constant" by order one is $\sim 10^{131}$y.  Of course, it could change  much faster, without 
 any conflict with  (\ref{central}). 
 
 In conclusion, all we can say with a relative rigor  is that eternally-constant positive cosmological constant  appears to be in conflict with basic principles of quantum gravity, but to restrict its variation from first principles, to an experimentally observable level, shall require some model building work.
 
 Let us finish this note on a highly speculative optimistic comment:
 What if we can extend cosmology beyond the point of quantum breaking? 
 Since, according to (\ref{central}), a generic meta-stable deSitter state faces 
 the quantum breaking before tunnelling, we could turn this 
 to our advantage and ask if the phenomenon of quantum breaking could 
 provide a new type of graceful exit for models, such as, 
Guth's original meta-stable inflationary scenario.   

\section*{Acknowledgements}
This work was supported in part by the Humboldt Foundation under Humboldt Professorship Award, ERC Advanced Grant 339169 "Selfcompletion", by TR 33 "The Dark Universe", and by the DFG cluster of excellence "Origin and Structure of the Universe".

\end{document}